\documentclass[a4paper,11pt]{article}

\usepackage{pos}

\title{One or two poles for the $\Xi(1820)$?}

\author*[a]{R.~Molina}
\author[b,c]{Wei-Hong Liang}
\author[b,c]{Chu-Wen Xiao}
\author[d,e]{Zhi-Feng Sun}
\author[a,b]{E.~Oset}

\affiliation[a]{Departamento de F\'{\i}sica Te\'orica and IFIC, Centro Mixto Universidad de
Valencia-CSIC Institutos de Investigaci\'on de Paterna, Aptdo.22085,
46071 Valencia, Spain}
\affiliation[b]{Department of Physics, Guangxi Normal University, Guilin 541004, China}

\affiliation[c]{Guangxi Key Laboratory of Nuclear Physics and Technology, Guangxi Normal University, Guilin 541004, China}

\affiliation[d]{Lanzhou Center for Theoretical Physics, Key Laboratory of Theoretical Physics of Gansu Province,
and Key Laboratory of Quantum Theory and Applications of MoE, Lanzhou University, Lanzhou, Gansu 730000, China}
\affiliation[e]{Research Center for Hadron and CSR Physics, Lanzhou University and Institute of Modern Physics of CAS, Lanzhou 730000, China}

\emailAdd{rmolina@ific.uv.es}

\abstract{In this talk, we present a new interpretation for the recently observed $\Xi(1820)$ resonance. We recall that the chiral unitary approach for the interaction of
pseudoscalar mesons with the baryons of the decuplet predicts two states
for the $\Xi(1820)$ resonance, one with a narrow width and the other one
with a large width. We contrast this fact with the recent BESIII
measurement of the $K^- \Lambda$ mass distribution in the $\psi(3686)$ decay to
$K^- \Lambda \bar\Xi^+ $, which demands a width much larger than the average
of the PDG, and show how the consideration of the two $\Xi(1820)$ states
provides a natural explanation to this apparent contradiction.}

\FullConference{%
  QNP 2024\\
  July 8, 12, 2024 \\
  Universitat de Barcelona, Barcelona, Spain 
}

\begin{document}
\maketitle

\section{Introduction}

We have now strong signatures that two pole structures indeed do exist. For that we mean two distinct poles in the same Riemann sheet. One of the most well-known examples is the $\Lambda(1405)$,
for which two states around $1385 \; \rm MeV$ and $1420 \;\rm MeV$ were predicted in Refs.~\cite{ollerulf,cola}. After some time, these resonances found their place in the PDG \cite{pdg}. Recently, the two-pole structure of the $\Lambda(1405)$ has also been observed by LQCD (for a pion mass around $200$)~\cite{BaryonScatteringBaSc:2023ori,BaryonScatteringBaSc:2023zvt}. The first chiral extrapolation to the physical point of these data has been conducted in~\cite{Zhuang:2024udv}. 
Other examples are the $K_1(1270)$ axial vector resonance,
where also two states were found in Ref.~\cite{rocasingh}, see also Ref.~\cite{gengroca}, and the $D^*(2400)$ \cite{solertwo,gamer}, the $Y(4260)$ resonance found by the BaBar collaboration \cite{BaBar1, BaBar2} into
two states $Y(4230)$ and $Y(4260)$ by the BESIII collaboration \cite{BES4230}. The generation of double pole structures from the Weinberg-Tomozawa interaction has been discussed in~\cite{GengZou}.

In Refs.~\cite{Weise,osetramos}, the interaction of the octet of pseudoscalar mesons with the octet of baryons was studied
and the two $\Lambda(1405)$ states emerged. The study was extended to the interaction of the octet of pseudoscalar mesons with the decuplet of baryons in  Ref.~\cite{Sarkar}. There, many resonances were generated that could be associated to well known $\frac{3}{2}^-$ existing states. Indeed, two resonances, one narrow and one with a large width, were predicted in Ref.~\cite{Sarkar} in the vicinity of $\Xi(1820)$.
The purpose of this work is to show that support for this idea is now provided by the recent BESIII investigation of this resonance.
Actually, in Ref.~\cite{BESshen} the $\psi(3686)$ decay to $K^- \Lambda \bar \Xi^+$ is investigated
and in the $K^- \Lambda$ invariant mass two neat peaks, one for the $\Xi(1690)$ and another one for the $\Xi(1820)$, are observed.
The surprising thing is that the width of the $\Xi(1820)$ is reported as
\begin{equation}\label{eq:widBES}
	\Gamma_{\Xi(1820)}=73^{+6}_{-5} \pm 9 \; \rm MeV.
\end{equation}
This result is much bigger, and incompatible with that of the PDG \cite{pdg} of
\begin{equation}\label{eq:widPDG}
	\Gamma_{\Xi(1820)}^{\rm PDG}=24^{+15}_{-10}  \; \rm MeV {\rm ~(PDG ~estimate)}; ~~~~24 \pm 5 \; \rm MeV ~{\rm (PDG ~average)}.
\end{equation}
A solution to this problem is obtained with the acceptance of two states, as we show below.

\section{Formalism}

In Ref.~\cite{Sarkar}, four coupled channels were considered, $\Sigma^* \bar K [1878], \Xi^* \pi [1669], \Xi^* \eta [2078]$ and $\Omega K [2165]$,
where the threshold masses are written in brackets in units of $\rm MeV$.
As one can see, only the $\Xi^* \pi$ channel is open for decaying at $1820 \; \rm MeV$ and the width of a state depends on the coupling to this channel.
The transition potential obtained from the chiral Lagrangians is given by
\begin{equation}\label{eq:Vij}
	V_{ij}=-\dfrac{1}{4f^2} C_{ij} (k^0+k^{\prime \,0}),
\end{equation}
where $k^0, k^{\prime \,0}$ are the energies of the initial and final mesons,
and the coefficients $C_{ij}$ are given in Table~1 of Ref.~\cite{Molina:2023uko}. The scattering amplitude can be evaluated with the Bethe-Salpeter (BS) equation,
\begin{equation}\label{eq:BS}
  T=[1-VG]^{-1} \, V.
\end{equation}
In this way, two poles were obtained in Ref.~\cite{Sarkar},
one narrow and the other one wide, in the vicinity of $\Xi(1820)$ resonance.

\begin{figure}[t]
\begin{center}
\includegraphics[scale=0.7]{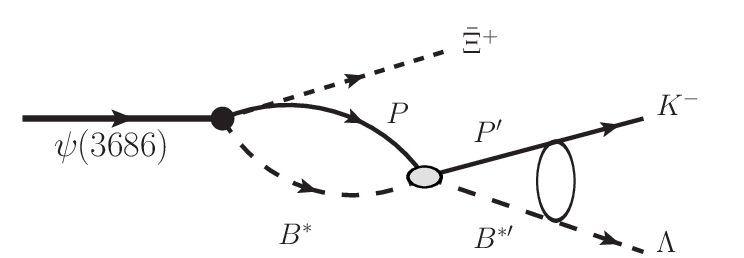}
\end{center}
\vspace{-0.7cm}
\caption{The resonant mechanism for the production of $\bar \Xi^+ K^- \Lambda$ in the $\psi(3686)$ decay.}
\label{Fig:Psidecay}
\end{figure}

Even though the channel $K^- \Lambda$, where the state is observed \cite{BESshen},
is not any of the coupled channels in the chiral unitary approach~\cite{Sarkar,Molina:2023uko}, there is a way to make a transition to this state by means of the mechanism depicted in Fig.~\ref{Fig:Psidecay},
which is of the type~\cite{Molina:2023uko},
\begin{eqnarray}\label{eq:t}
  t&=& \sum_j A_j \, \vec \epsilon_\psi \cdot \vec p_{\bar \Xi} \, G_j (PB^*) \, T_{ji} \, C_i \, \tilde{k}^2   
  \sim \sum_{ij} D_{ij}\, \tilde{k}^2 \, \vec \epsilon_\psi \cdot \vec p_{\bar \Xi} \, T_{ji},
\end{eqnarray}
where $\tilde{k}$ is the momentum of the $K^-$ in the $K^- \Lambda$ rest frame,
$G_j$ are the loop functions of the intermediate $PB^*$ states,
regularized by means of a cutoff $q_{\rm max}$ \cite{osetramos},
and $A_j, C_i, D_{ij}$ are unknown coefficients that depend on the dynamics in Fig.~\ref{Fig:Psidecay}. The invariant mass distribution can be written as,
\begin{equation}
\frac{{\rm d}\Gamma}{{\rm d}M_{\rm inv}(K^- \Lambda)} = \frac{1}{(2\pi)^3} \;\frac{1}{4M^2_{\psi}} \; p_{\bar{\Xi}} \,\tilde{k} \;\bar{\sum} \sum |t|^2,
\end{equation}
where $p_{\bar{\Xi}}$ is the momentum of the $\bar{\Xi}$ in the $\psi(3686)$ rest frame,
and $\tilde{k}$ is the momentum of the kaon in the c.m. reference system of the $K^-\Lambda$, $\tilde{k}=\lambda^{1/2}(M_\mathrm{inv}^2, m_K^2, m_{\Lambda}^2)/2 M_\mathrm{inv}$.
We obtain
\begin{equation}
\frac{{\rm d}\Gamma}{{\rm d}M_{\rm inv}(K^- \Lambda)} = W \, p^3_{\bar{\Xi}}\; \tilde{k}^5 \;\sum_{ij}  \left| D_{ij} \,T_{ji} \right|^2,
\label{eq:dgainv}
\end{equation}
being $W$ an arbitrary weight.
\section{Results}
In the first place, we have redone here the calculations of Ref.~\cite{Sarkar} corroborating them. We have also checked that the results presented here
are also stable by varying the parameters ($f$ and $q_\mathrm{max}$). We obtain two-poles,
one with a small width and the other one broad.
The best compromise with the experimental data is obtained by slightly changing the $f$ parameter in Eq.~\eqref{eq:Vij} to $1.28 f_\pi$, and $q_\mathrm{max}=830$ MeV.
The results are shown in Table~\ref{tab:result},
together with the couplings of the states to the different channels,
extracted from the behaviour at the pole, where the amplitude behaves like $T_{ij} \simeq g_i g_j /(\sqrt{s} - M_R)$.
It is now clear why the two states have such a different width,
since the only decay channel is $\pi \Xi^*$ and the width goes as the square of the coupling to that channel,
which is larger for the second state.
\begin{table}[tb]
     \renewcommand{\arraystretch}{1.2}
     \setlength{\tabcolsep}{0.3cm}
\centering
\caption{ Pole positions and couplings for $q_{\rm max} = 830$ MeV. All quantities are given in units of MeV.}
\label{tab:result}
\begin{tabular}{c|c|c|c}
\hline
Poles  & $|g_i|$ & $g_i$ & channels  \\
\hline
$1824 - 31 i$ & 3.22 & $3.22 - 0.096 i$  & $\bar{K} \Sigma^*$  \\
                       & 1.71 & $1.55 + 0.73 i$ & $\pi \Xi^*$  \\
                       & 2.61 & $2.58 - 0.38 i$    & $\eta \Xi^*$  \\
					   & 1.62 & $1.47 + 0.67 i$   & $K \Omega$  \\
\hline
$1875 - 130 i$ & 2.13 & $0.29 + 2.11 i$    & $\bar{K} \Sigma^*$  \\
                         & 3.04 & $-2.07 + 2.23 i$  & $\pi \Xi^*$  \\
                         & 2.20 & $1.11 + 1.90 i$    & $\eta \Xi^*$  \\
					     & 3.03 & $-1.77 + 2.45 i$   & $K \Omega$  \\
\hline\hline
\end{tabular}
\end{table}

\begin{figure}[ht]
\centering
\includegraphics[scale=0.75]{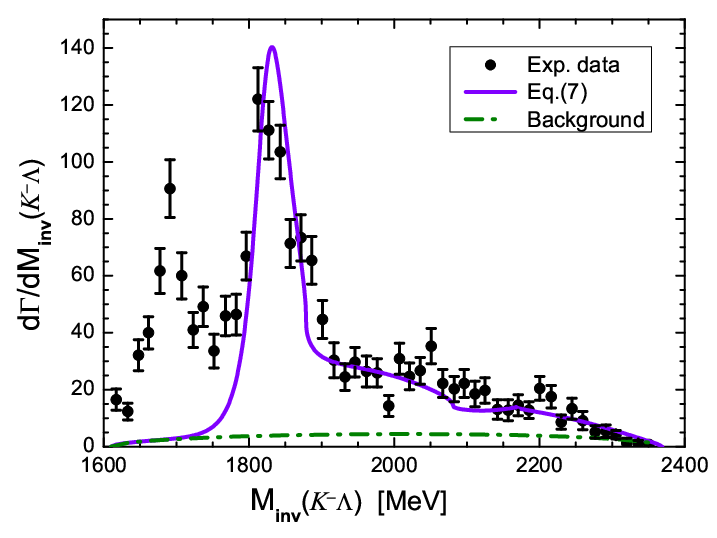}
\caption{Results of Eq.~\eqref{eq:dgainv}, in arbitrary units, with $\sum_{ij} D_{ij} T_{ji}$ substituted by $T_{33}$, with the experimental data taken from BESIII \cite{BESshen} and the background as explained in the text.}
\label{fig:massdis}
\end{figure}

Secondly, we have evaluated the mechanism of Fig.~\ref{Fig:Psidecay} and compared with the experimental data. The coefficients $D_{ij}$ in Eq.~(\ref{eq:dgainv}) are unknown.
However, by looking at the strength of the different $T_{ij}$ matrices,
we find that the $\eta \Xi^*$ channel has a large diagonal $T_{33}$ amplitude
which shows evidence of the broad resonance (this is in agreement with Fig. 7 of Ref.~\cite{Sarkar}).
We take then this amplitude characterizing the sum $\sum_{ij} D_{ij} T_{ji}$.
Actually, we notice that the relevant $T_{ij}$ matrix elements have all a similar shape. Once this is done,
we find a $K^- \Lambda$ mass distribution as shown in Fig.~\ref{fig:massdis}. We add a background that follows the phase space, $
C \;p_{\bar{\Xi}} \;\tilde{k} $,
and adjust $C$ and the strength to the experimental data. 

\begin{figure}[t]
\centering
\includegraphics[scale=0.75]{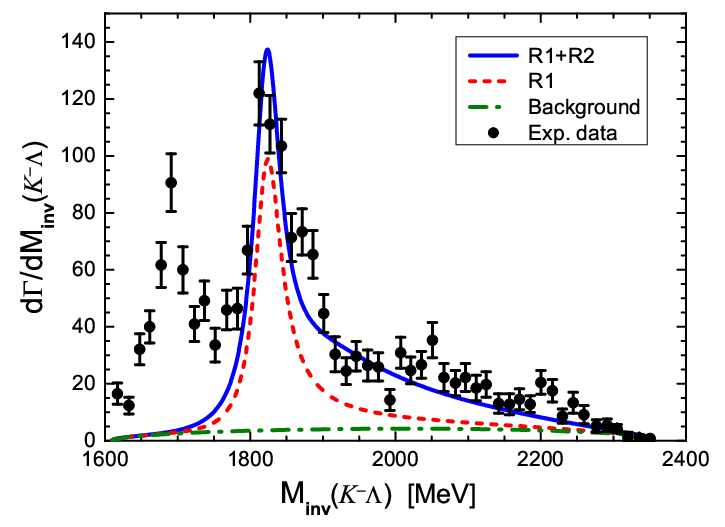}
\caption{Results obtained adjusting Eq.~\eqref{eq:breit} to the data together with a small background. In the figure we show the contribution of the background alone and the results obtained removing the contribution of the second pole.}
\label{fig:massdis2}
\end{figure}

As we can see, the results obtained with the two resonances of Table~\ref{tab:result},
together with the background, provide a fair description of the data.
A second test is conducted by performing a fit to the data, very similarly to what is usually done in experimental analyses. Thus, we take a coherent sum of amplitudes
\begin{equation}
 \frac{A}{M_{\rm inv} - M_{R_1} + i \frac{\Gamma_1}{2}} + \frac{B}{M_{\rm inv} - M_{R_2} + i \frac{\Gamma_2}{2}},
\label{eq:breit}
\end{equation}
with $R_1,\; R_2$ representing approximately the two resonances of Table~\ref{tab:result},
with $M_{R_1} = 1822$ MeV, $\Gamma_1 = 45$ MeV, $M_{R_2} = 1870$ MeV, $\Gamma_2 = 200$ MeV.
We adjust $A$ and $B$ and the constant $C$ for the background.
The coefficients $A$ and $B$ are found to have about the same strength.
A good description of the data, shown in Fig.~\ref{fig:massdis2}, is obtained,
and most of the strength at higher invariant masses is provided by the contribution of the second resonance.
We can see in Fig.~\ref{fig:massdis2} that the contribution of the wider resonance plays an important role
filling up the strength in the higher part of the mass spectrum. Note that the background needed in Figs. \ref{fig:massdis} and \ref{fig:massdis2} is practically the same.
This means that in Fig. \ref{fig:massdis} the upper part of the spectrum comes from the $T_{33}$ amplitude,
which contains information of the two resonances, with the wide one responsible for the strength in this region.

Note also that around $M_{K^- \Lambda} = 2100$ MeV there appears another peak also obtained in Ref.~\cite{Sarkar}. This peak is  more related to the $K \Omega$ channel.

In conclusion, the chiral unitary approach for meson-baryon interaction,
applied to the interaction of pseudoscalar mesons with the baryon-decuplet, gives rise to two states around the $\Xi(1820)$,
one of them narrow and the other one wide.
This feature remains if reasonable changes are done in the strength of the interaction or the regulator of the loop functions,
and is independent on whether one uses dimensional regularization \cite{Sarkar} or the cutoff method as done here. We have shown that this scenario provides a satisfactory description of the experimental data in the $\psi(3686) \to K^- \Lambda \bar{\Xi}^+$ decay,
solving the puzzle presented by the recent BESIII experiment \cite{BESshen}, which provides a larger width for the $\Xi(1820)$ than the one reported in the PDG \cite{pdg}.

\section*{ACKNOWLEDGEMENT}
This work is partly supported by the National Natural Science Foundation of China under Grant No. 11975083 and No. 12365019, and by the Central Government Guidance Funds for Local Scientific and Technological Development, China (No. Guike ZY22096024).
R. M. acknowledges support from the CIDEGENT program with Ref. CIDEGENT/2019/015,
the Spanish Ministerio de Economia y Competitividad
and  European Union (NextGenerationEU/PRTR) by the grant with Ref. CNS2022-13614.
This work is also partly supported by the Spanish Ministerio de Economia y Competitividad (MINECO) and European FEDER
funds under Contracts No. FIS2017-84038-C2-1-P B, PID2020-112777GB-I00, and by Generalitat Valenciana under contract
PROMETEO/2020/023.
This project has received funding from the European Union Horizon 2020 research and innovation
programme under the program H2020-INFRAIA-2018-1, grant agreement No. 824093 of the STRONG-2020 project.

\end{document}